\documentclass[useAMS,usenatbib]{mn2e}
\usepackage{amsmath,amssymb}
\usepackage{natbib}
\usepackage{graphicx}
\usepackage{times}
\usepackage{threeparttable}

\def\ifnote{\iffalse}

\title[GRB 080319B high energy]{{The possible} high energy emission from GRB 080319B and
origins of the GeV emission of GRBs 080514B, 080916C and 081024B}
\author[Y. C. Zou, Y. Z. Fan \& T. Piran]{Yuan-Chuan Zou$^{1,3}$, Yi-Zhong Fan$^{2,4}$,
and Tsvi Piran$^{1}$ \thanks{Email: yizhong@nbi.dk (YZF) and
tsvi@phys.huji.ac.il (TP)}
\\
$^{1}${The Racah Institute of Physics, Hebrew University, Jerusalem 91904, Israel}\\
$^{2}${Niels Bohr International Academy, Niels Bohr
Institute, University of Copenhagen, Blegdamsvej 17, DK-2100
Copenhagen, Denmark} \\
$^{3}${School of Physics,
Huazhong University of Science and Technology,  Wuhan 430074, China}\\
$^{4}${Purple Mountain Observatory, Chinese Academy of
Sciences, Nanjing 210008, China}
}

\begin{document}
\date{\today}
\maketitle
\label{firstpage}

\begin{abstract}
We calculate the high energy (sub-GeV to TeV) prompt and afterglow
emission of GRB 080319B that was distinguished by a naked-eye
optical flash and by an unusual strong early X-ray afterglow. There
are three possible sources for high energy emission: the prompt
optical and $\gamma$-ray photons IC scattered by the accelerated
electrons, the prompt photons IC scattered by the early external
reverse-forward shock electrons, and the higher band of the
synchrotron and the synchrotron self-Compton emission of the
external shock. There should have been in total {hundreds} high
energy photons detectable for the Large Area Telescope (LAT) onboard
the Fermi satellite, and {tens} photons of those with energy $> 10$
GeV. The $> 10$ GeV emission had a duration about twice that of the
soft $\gamma$-rays. AGILE could have observed these energetic
signals if it was not occulted by the Earth at that moment. The
physical origins of the high energy emission detected in GRB
080514B, GRB 080916C and GRB 081024B are also discussed. These
observations {seem to be consistent with the current high energy
emission models}.
\end{abstract}
\begin{keywords}
 gamma rays: bursts$-$radiation mechanism: nonthermal
\end{keywords}

\section{Introduction}
\begin{figure}
\includegraphics{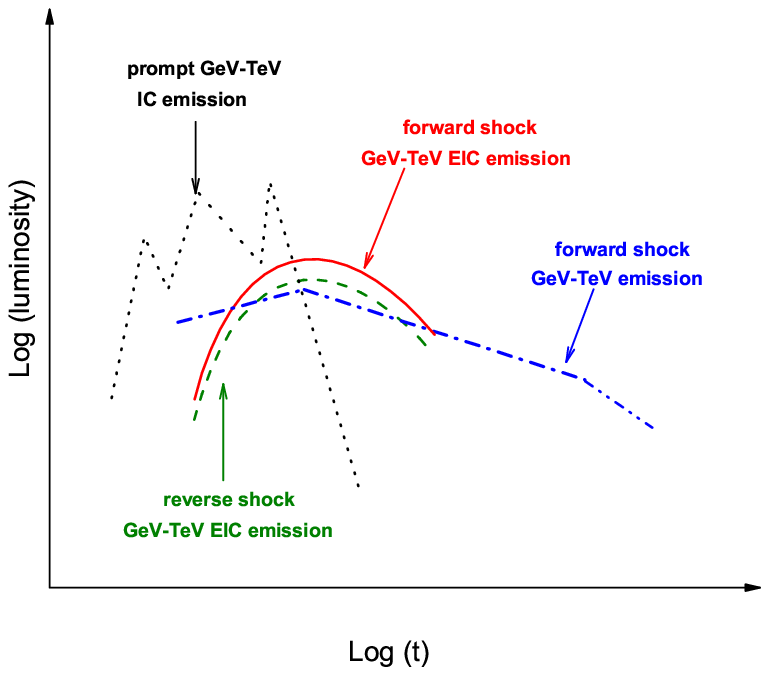}
\caption{Schematic light curves for the different component of the
high energy emissions: {prompt} SSC, reverse shock EIC, forward
shock EIC and external shock SSC respectively.} \label{fig:sketch}
\end{figure}

A breakthrough of GRB observation, made by {\it Swift} satellite in
2008, is the discovery of the very bright burst GRB 080319B which
was accompanied by a naked-eye optical flash \citep{Racusin08b}. The
optical observation was going on even before the onset of the
$\gamma$-ray burst because TORTORA was monitoring the same region of
the sky at that moment \citep{Cwiok08, Karpov08}. The X-ray
telescope (XRT) onboard {\it Swift} satellite slewed to the source
about $60$ sec after the trigger of the burst and recorded a quickly
decaying but extremely bright X-ray component. These continuous
observations collected fruitful data \citep{Racusin08b,Bloom08} and
rendered GRB 080319B one of the best-studied bursts so far. Although
no very high-energy emission was directly detected from GRB 080319B
the unique spectrum of this burst and its afterglow suggest that it
has been accompanied by a very strong GeV-TeV emission that would
have already been detected by AGILE if not occulted by earth at that
moment. Based on a model in which the prompt optical and soft
$\gamma$-ray emission are respectively the synchrotron and the first
order inverse Compton (IC) radiation components of the internal
shocks, Kumar \& Panaitescu (2008), Racusin et al. (2008) and Fan \&
Piran (2008) suggested that the second order IC of the internal
shocks would peak in GeV-TeV energy range and the isotropic energy
might be high up to $\sim 10^{55}$ erg (see however Piran, Sari \&
Zou 2008 and Fan, Zhang \& Wei 2009). Because of the tight
overlapping of the prompt emission with the reverse/forward shock
regions, some soft $\gamma$-rays will be up-scattered by the reverse
shock electrons and some prompt optical photons will be up-scattered
by the forward shock electrons, i.e., the so-called external inverse
Compton (EIC). As a result, two additional GeV-TeV emission
components with a duration $\sim 100$ s are expected (Fan \& Piran
2008). In this work, we discuss these possibilities in more detail.
Moreover, we show that the early ($60-2000$ s) forward shock
synchrotron and the synchrotron self-Compton (SSC) emission in the
energy range $20{\rm MeV}-300{\rm GeV}$ is as powerful as the high
energy emission detected in GRB 080916C \citep{Tajima08}. A
schematic plot of the expected GeV-TeV signals from GRB 080319B is
shown in Fig.\ref{fig:sketch}.

Since its successful launch on June 11 2008, the Fermi satellite has
detected the prompt $>10$ GeV emission in GRB 080916C
\citep{Tajima08,Omodei08b}, and the GeV emission following a short
burst GRB 081024B \citep{Omodei08}. As GRB 080514B
\citep{Giuliani08}, GRB 080825C \citep{Bouvier08}, and some other
events detected by the Compton Gamma Ray Observatory (CGRO)
satellite in 1991-2000 \citep{Hurley94,Gonz03}, the high energy
emission of both GRB 080916C and GRB 081024B lasted longer than the
prompt soft $\gamma$-rays. The detection of high energy signals
sheds some lights on the bulk Lorentz factor of the ejecta, the
radiation mechanisms, the physical composition of the outflow and
the prolonged activity of the central engine. This is particularly
the case if the simultaneous X-ray/optical emission data are
available (see Fan \& Piran 2008 for a recent review). In this work
we'll outline the origins of the GeV emission from GRB 080514B, GRB
080916C and GRB 081024B, based on the (preliminary) public data.

The paper is structured as follows. In Section 2, we calculate the
possible prompt and afterglow GeV-TeV emission of GRB 080319B. In
section 3, we interpret the high energy emission detected in GRB
080514B, GRB 080916C and GRB 081024B. In section 4, we summarize our
results with some discussions.

\section{Possible GeV-TeV emission from GRB 080319B}
GRB 080319B \citep{Racusin08b} was most notable due to its huge
total energy and especially its extremely luminous prompt optical
emission that could be seen with naked eyes \citep{Cwiok08,
Karpov08}. This burst was located at a redshift $z=0.937$ space
\citep{Vreeswijk08} and duration was $T_{90}\sim $57s. The peak
energy of the {$\nu F_{\nu}$ spectrum} was $E_{\rm p} \simeq
675\pm22$ keV, and the photon indexes below and above $E_p$ were
$-0.855^{+0.014}_{-0.013}$ and $-3.59^{+0.32}_{-0.62}$ respectively.
Choosing standard cosmological parameters $H_0=70 {\rm km}\,{\rm
s}^{-1}\,{\rm Mpc}^{-1}, \Omega_{\rm M}=0.3,
\Omega_\Lambda=0.7$(corresponding to a luminosity distance $D_L \sim
1.9\times 10^{28}$cm), we have a peak luminosity $L_{\rm peak} \sim
1.0\times 10^{53} {\rm erg}\,{s}^{-1}$ and an isotropic energy
$E_{\rm iso} \simeq 1.3 \times 10^{54}$ erg
\citep{Racusin08b,Bloom08,ga08}. \citet{Karpov08} reported the
optical V-band ($\sim 6 \times 10^{14}$Hz) light curve in the prompt
phase (from $\sim$ -10 s to $\sim$ 100 s). Variability was evident
and there were at least 3 or 4 main pulses in the light curve. The
peak V-band reached magnitude of 5.3, corresponding to a flux
density $\sim 28.7$ Jy, and {isotropic equivalent} energy $E_{opt}
\sim 2 \times 10^{52}$erg if we take $\sim 20$ Jy as the average
flux density. The variability and the very sharp decline of the
prompt optical emission support an internal origin of these optical
photons, though the underlying physical process is not clear yet
\citep[see][for a discussion of various possible models]{zps08}.

Afterglow modeling can in principle constrain the total kinetic
energy and the initial Lorentz factor of the GRB ejecta, and the
physical parameters of the external shocks (Sari, Piran \& Narayan
1998; Chevalier \& Li 2000; Panaitescu \& Kumar 2001). The behavior
of the afterglow of GRB 080319B suggests a free wind medium
\citep{kp08,Racusin08,Wu08}. A self-consistent modeling of the X-ray
and optical afterglow data favors a two-component jet model
\citep{Racusin08,Wu08}. Moreover, the shock parameters of the narrow
and wide ejecta components need to be very different, as found in
GRB 051221A \citep{Jin07}. Following \citet{Racusin08} and
\citet{Wu08}, we take the isotropic kinetic energy of the narrow
ejecta (represented by the subscript $``{\rm n}"$) $E_{\rm k,n}\sim
3\times 10^{55}$ erg\footnote{An $E_{\rm k,n}$ high up to $\sim
10^{55}$ erg is rather unusual. Similar result has only been
reported in the afterglow modeling of GRB 060418 \citep{JF07}.
However we believe that such a huge value is possible for GRB
0980319B because the XRT flux at $t\sim 70$ s is as bright as $\sim
10^{-7}~{\rm erg~s^{-1}~cm^{-2}}$, which is the brightest X-ray
afterglow detected so far and is even much brighter than most prompt
X-ray emission of {\it Swift} GRBs. On the other hand both the
spectral and the temporal behaviors of the early ($60-2000$ s) X-ray
emission strongly favor a fireball model in the slow cooling phase,
which requires small $\epsilon_{\rm B,n}$ and $A_*$. As a result, we
do need an $E_{\rm k,n} \sim 10^{55}$ erg to reproduce the
observation data (see footnote \ref{foot:2}).}, the wind parameter
$A_* \sim 0.01$, the fraction of forward shock energy given to the
electrons $\epsilon_{\rm e,n}\sim 0.1$, the fraction of forward
shock energy given to the magnetic field\footnote{We do not take
$\epsilon_{\rm n,B} \sim 10^{-6}$ as in \citet{Racusin08} (see
section \ref{sec:EIC} below) because the peak flux density of the
forward shock synchrotron emission is $F_{\nu,\max}^{\rm syn} \sim 9
 \epsilon_{\rm B,n,-4}^{1/2} E_{\rm k,n,55.5}^{1/2} A_{\star,-2} D_{L,28.3}^{-2}  t_3^{-1/2}
{\rm mJy}$. At $t\sim 60$ sec, the X-ray (at 1 keV) flux $\sim 20$
mJy \citep{Bloom08} disfavors an $\epsilon_{\rm B,n}$ as small as
$\sim 10^{-6}$. On the other hand, an $\epsilon_{\rm B,n} \sim
10^{-6}$ will give rise to a too large cooling Lorentz factor
$\gamma_c \sim 10^{10}(1+Y_{\rm ssc})^{-1}$, where the forward shock
SSC parameter  $Y_{\rm ssc}\ll \sqrt{\epsilon_{\rm
n,e}/\epsilon_{\rm B,n}}$ since the SSC emission of such energetic
electrons should be in Klein-Nishina regime and thus be effectively
suppressed.\label{foot:2}} $\epsilon_{\rm B,n}\sim 10^{-4}$, the
power-law distribution index $p_{\rm n} \sim 2.4$, and the
half-opening angle $\theta_{\rm j,n} \sim 0.2$ degree. We do not
discuss the wide jet component because it plays a less important
role in producing GeV-TeV afterglow emission. The average Lorentz
factor of the narrow jet outflow ($\Gamma_{\rm i}$) before getting
decelerated by a stellar wind medium is very high. A lower limit can
be set by the Lorentz factor of the forward shock at $\sim 70$ s,
when the X-ray afterglow began to decline normally, i.e.
\citep{BM76,DL98},
\[
\Gamma \simeq 600 E_{\rm
k,n,55.5}^{1/4}A_{*,-2}^{-1/4}(t/70{\rm s})^{-1/4}[(1+z)/2]^{1/4}.
\]
So a choice of $\Gamma_{\rm i}\sim 1000$ is rather reasonable.
Throughout this work we adopt the convenience $Q_{\rm x}=Q/10^{\rm
x}$ in units of cgs

In the leading fireball model for GRBs \citep[see][for
reviews]{Piran04,Mesz02,Zhang07}, the synchrotron and IC radiation
will give rise to a high-energy component that will be emitted along
with the prompt sub-MeV photons and the afterglow
radio/optical/X-ray emission \citep{fp08}.
Depending on the seed photons' origins, IC can be
SSC or EIC. Below we'll show that for GRB
080319B both processes plausibly played an important role in
producing GeV$-$TeV emission. This suggests that similar bursts will
provide promising sources for the Fermi high energy satellite.

\subsection{Prompt GeV-TeV IC emission}
\citet{zps08} showed that the SSC models in which the soft
$\gamma$-rays are the IC component of the optical photons cannot
explain the observations. The major obstacle is the resulting high
synchrotron self-absorption frequency and then the X-ray spectrum
that is inconsistent with the observation. If we ignore this
problem, there is a solution with a Compton parameter $Y \sim 1$ and
a stochastic Lorentz factor $\gamma_{\rm e} \sim 100$. Then the 2nd
IC peaks at $2 \gamma_{\rm e}^2 E_{\rm p} \sim 15$ GeV, and the
number of the detectable photons is $Y E_\gamma S_{\rm det}/4\pi
D_L^2 h \nu_{\rm 2nd,IC}$ corresponding to detected $\sim 130$
photons by LAT, with $S_{\rm det} \sim 10^4 {\rm cm}^2$ at GeV
energies. Other models with larger $Y$ lead to even stronger signals
\citep{kp08,Racusin08,fp08}.

{As discussed in \citet{zps08}, too much energy should be hidden
either in the high energy component if using the SSC model, or in
the low energy component of electrons if we assume the prompt
optical emission and $\gamma$-rays were from synchrotron emission by
two components of electrons in the same region. It indicates that
the two different bands of prompt photons should come from different
geometrical regions.} Below we consider these two different regions
{(possibly but not necessarily two sets of different internal
shocks) within the same outflow cone}, denoted by the subscripts
``${\rm opt}$" and ``$\gamma$" respectively. Strong high energy
prompt emission is still possible, and it can be estimated even
though the {details of the internal shocks are} still unclear. The
possible high energy emission consists of four components: self-IC
scattering in the optical emission region; self-IC scattering in the
$\gamma$-ray producing region; optical photons IC scattered in the
$\gamma$-ray producing region; and soft $\gamma$-rays IC scattered
in the optical emission region. Note that because of the steep
decline of the high energy slope ($\beta \sim 2.6$) extrapolation of
the soft $\gamma$-ray emission gives only a very weak signal.

\subsubsection{Prompt optical emission region}
The observed optical flux density limits the temperature of this
region (see the Appendix for the derivation):
\begin{equation}
f_{\rm \nu,opt} \leq \frac{2\pi \nu^2_{\rm opt} (1+z)^3 \Gamma_{\rm
opt} k T_{\rm opt}}{c^2} \left(\frac{R_{\rm opt}}{\Gamma_{\rm opt}
D_L}\right)^2,\label{eq:main1}
\end{equation}
where $\Gamma_{\rm opt}$ is the bulk Lorentz factor, $k$
is Boltzmann constant, $T_{\rm
opt}$ is the temperature (while the minimal temperature $T_{\rm
opt,\min}$ corresponds to the equality), and $R_{\rm opt} $ is the
emission region radius. Thus
\begin{equation}
k T_{\rm opt,\min} = 6 \times 10^{-5} \Gamma_{\rm opt,3} R_{\rm
opt,16}^{-2}~ {\rm erg}.
\end{equation}
Noticing that the bulk Lorentz factor in the afterglow is
high \citep{Racusin08}, we take a fiducial  value of
$\Gamma_{opt}\sim 10^3$ for the
prompt phase. Considering the variability of the light curves
and the deceleration radius, which constrains the radius should not
be too large, then the choice $10^{16}$cm is reasonable. The
corresponding typical stochastic Lorentz factor of the electrons is
\begin{equation}
\gamma_{\rm e,opt,\min} \sim k T_{\rm opt,\min}/(m_{\rm e} c^2) \sim
75 \Gamma_{\rm opt,3} R_{\rm opt,16}^{-2}, \label{eq:gamma_e,min}
\end{equation}
where $m_{\rm e}$ is the rest mass of the electron.

The first order IC is in the soft $\gamma$-ray band. As mentioned
before the prompt soft $\gamma$-rays are unlikely to be the first
order IC component of the optical emission. So the first order IC
radiation of the electrons emitting optical photons would be much
smaller than the detected soft $\gamma$-rays.
Correspondingly, the 2nd order IC radiation in GeV-TeV energy range
is unimportant as it falls below the IC radiation that arises when the prompt soft
$\gamma$-rays cross the optical emission region.

\subsubsection{$\gamma$-rays IC scattered in the prompt optical emission region}
\label{sec:gamma-opt}
If the soft $\gamma$-rays pass through the prompt optical
emitting electrons, the ``optical depth" for electrons is approximately
$
 \sigma_T \frac{N_{\gamma} \delta t_{\rm opt}/T_{90}}{4\pi R^2}
{\ifnote
= 6.65e-25  \frac{1.3e60 }{4\pi 1e32 R_{16}^2} \\
\fi} \sim 3 N_{\gamma,60}R_{16}^{-2} \delta t_{\rm opt,-0.5}, $
 where $\delta t_{\rm opt} \sim 0.3~R_{\rm
opt,16}\Gamma_{\rm opt,3}^{-2}$s is the typical variability
timescale of the prompt optical emission. For each collision the
electron loses energy $\sim \gamma_{\rm e,opt}^2
h\nu_{\gamma}/\Gamma_{\rm opt} < \gamma_{\rm e,opt} m_e c^2 $ as
long as $\gamma_{\rm e,opt} < \Gamma_{\rm opt}$.

Assuming that almost all electrons carried by the GRB outflow
contributed to the prompt optical emission, which should be an upper limit,
we estimate the
number of electrons that participate in a typical optical pulse (with a
variability timescale $\delta t_{\rm opt}$):
\begin{equation}
N_{\rm e,p,opt} \sim  {E_{\rm k,n}\delta t_{\rm opt}\over
\Gamma_{\rm i}m_{\rm p}c^2T_{90}}
 \sim 10^{53}E_{\rm k,n,55.5}\delta t_{\rm opt,-0.5}\Gamma_{\rm
 i,3}^{-1}.
 \label{eq:N_eopt}
\end{equation}

Using this value we estimate the optical depth for soft
$\gamma-$rays being scattered by the
electrons emitting the prompt optical emission as $\tau \sim
\sigma_{\rm T}N_{\rm e,p,opt}/(4\pi R_{\rm opt}^2)\sim 5 \times
10^{-5}E_{\rm k,n,55.5}\delta t_{\rm opt,-0.5}\Gamma_{\rm
i,3}^{-1}R_{\rm opt,16}^{-2}$. The total number of the IC photons
detectable by LAT is thus
\begin{equation}
N_{\rm det,\gamma-\rm opt} \sim {\tau N_{\gamma} S_{\rm det} \over
4\pi d_L^2} \leq  100 E_{\rm k,n,55.5}{\delta t_{\rm
opt,-0.5}}\Gamma_{\rm i,3}^{-1}R_{\rm opt,16}^{-2}N_{\gamma,60},
\label{eq:opt-gamma}
\end{equation}
where $N_{\gamma}$ is the total number of prompt soft $\gamma-$rays.
The typical energy of the IC photons
is greater than $E_{\rm IC,\gamma-opt}\sim 2\gamma_{\rm
e,opt,\min}^2 E_p \sim 8 \Gamma_{\rm opt,3}^2 R_{\rm opt,16}^{-4}$
GeV. The corresponding total energy of these photons is
$\sim 5 \times 10^{53}$ergs.


\subsubsection{Soft $\gamma$-ray emission region}
Since there may be no suitable IC model for the soft $\gamma$-rays,
we assume that these soft $\gamma$-rays are the synchrotron emission
 at a radius $R_{\rm \gamma}$. To match the
peculiar spectrum of the soft $\gamma$-rays, the cooling Lorentz
factor $\gamma_{\rm c} \sim (1+z) \frac{6\pi m_{\rm e}
c}{\sigma_{\rm T} \Gamma B^2 \delta t_{\gamma}}$ should be
comparable to the typical Lorentz factor of the electrons
$\gamma_{\rm m}$ \citep{zps08}. $\sigma_{\rm T}$ is the Thompson's cross
section and $\delta t_{\gamma} \sim 0.1$s \citep{Margutti08}
is the variability timescale of the
soft $\gamma$-rays. The condition $E_{\rm p}\sim \Gamma_{\gamma} 2
\gamma_{e,\gamma}^2 \frac{q_{\rm e} B_\gamma}{2 \pi m_{\rm e} c} /(1+z)$
gives
\begin{equation}
B_\gamma \sim 16 \Gamma_{\gamma,3}^{-1/3} \delta t_{\gamma,
-1}^{-2/3} {\rm Gauss},
\end{equation}
where $q_{\rm e}$ is the electron's charge. The typical Lorentz
factor of the emitting electrons is thus
\begin{equation}
\gamma_{\rm e,\gamma} \sim 6 \times 10^4 \Gamma_{\gamma,3}^{-1/3}
\delta t_{\gamma,-1}^{1/3}.\label{eq:gamma_egamma}
\end{equation}
{This value is relatively too high for internal shocks. However,
here we don't need it come from the internal shocks necessarily. The
other energy dissipation mechanisms may produce high $\gamma_e$.}
The SSC will be deep in the Klein-Nishina regime, and pair avalanche
effect might exist \citep{psz08}, additional component of high
energy photons would peak at energy $\Gamma_\gamma^2 (m_e
c^2)^2/(h\nu_\gamma) \sim 400 \Gamma_{\gamma,3}^2$ GeV, where $h$ is
the Plank constant.

Using $f_{\rm \nu,\max} = (1+z) N_{\rm
e,\gamma} \Gamma_{\rm \gamma} m_{\rm e} c^2 \frac{\sigma_{\rm T}
B}{3 q_{\rm e} 4\pi D_L^2}$, we get the number of electrons for each pulse
$N_{\rm e,p,\gamma} \sim 9 \times 10^{49} \Gamma_{\gamma,3}^{-2/3}
\delta t_{\gamma,-1}^{2/3}$, and the total number of electrons
is then $N_{\rm e,\gamma} \sim N_{\rm e,p,\gamma} T_{90}/\delta t_\gamma
\sim 5 \times 10^{52}\Gamma_{\gamma,3}^{-2/3} \delta t_{\gamma,-1}^{-1/3}$.

The corresponding optical depth for Thompson scattering is $\tau \sim
\sigma_T N_{\rm e,\gamma}/(4\pi R_\gamma^2) \sim 5 \times 10^{-8}
N_{\rm e,p,\gamma,50}R_{\gamma,16}^{-2}$.
\footnote{By the time a single photon passes through a sub-shell, this sub-shell
expands by a factor of $\sim$2 in radius. Therefore subsequent scattering in
other sub-shells will be negligible and when considering the optical depth
a single sub-shell should be taken into account.}
 And the Compton parameter in
KN regime is $Y \sim \gamma_{e,\gamma}^2 \tau /[\gamma_{e,\gamma}
h\nu_\gamma/(\Gamma m_e c^2)]^2 \sim 0.03  \Gamma_{\rm \gamma,3}^2
N_{\rm e,p,\gamma,50}R_{\gamma, 16}^{-2}$\citep{psz08}.
The total energy of the avalanche loaded pairs is in the order
of $2 Y E_\gamma$ even all the first produced very high energy photons
are cooled into steady pairs.  The number of
detectable photons by LAT is then
$\sim 0.1 R_{\gamma,16}^{-2}$. It is thus undetectable
even without taking into account the large optical depth ($\sim 10$)
of the universe to such energetic photons (Stecker et al. 2006).

\subsubsection{optical photons IC scattered in $\gamma$-rays region}
If the optical photons are produced in smaller radii than the soft $\gamma$-rays (
$R_\gamma \geq R_{\rm opt}$), they would be
 IC scattered in the $\gamma$-rays region. In this case, the electrons will be
 cooled to a random Lorentz factor $\gamma_{\rm e,\gamma,c} < 1.8\times 10^{4}
 \Gamma_{\rm \gamma,3}^3R_{\gamma,16}L_{\rm opt,50.7}^{-1}<\gamma_{\rm e,\gamma}$ \citep{fp08},  where
 $L_{\rm opt}>5\times 10^{50}{\rm erg~s^{-1}}$ is the luminosity
 of the prompt optical emission,
 suggesting that all the electrons were cooled by the IC scattering.
 The typical energy of the IC scattered photons is
$E_{\rm IC,opt-\gamma}\sim 2\gamma_{\rm
e,\gamma}^2 h \nu_{opt} \sim 14 \Gamma_{\gamma,3}^{-2/3} \delta t_{\gamma,-1}^{2/3}$GeV.
Since the electrons lost almost all the energy, the number of the
detectable photons by LAT is
\begin{equation}
N_{\rm det,opt-\gamma} \sim {E_{\rm e,\gamma}S_{\rm det}\over 4\pi
D_L^2 E_{\rm IC,opt-\gamma}} \sim 240 \Gamma_{\gamma,3}^{-1/3}
\delta t_{\gamma,-1}^{-2/3}
\end{equation}
where $E_{\rm e,\gamma}\approx \Gamma_{\gamma}\gamma_{\rm
e,\gamma}N_{\rm e,\gamma}m_{\rm e}c^2$ is the total energy carried
by the electrons emitting soft $\gamma-$rays.

{\it This discussion is valid only for $R_{\rm \gamma}\geq R_{\rm
opt}$ that is less likely}. As long as $R_{\rm opt} \gtrsim {\rm a
~few \times} R_{\rm \gamma}$, the prompt optical emission cannot
cool the accelerated electrons emitting soft $\gamma-$rays,
because the photons from $R_{\rm opt}$ reached $R_\gamma$ in a time
$\sim (R_{\rm opt}-R_{\rm \gamma})/c \sim 3\times 10^{5}~R_{\rm
opt,16}~{\rm sec}$ when the photons at $R_\gamma$ had been
disappeared long before. For the same reason,  there would be no
high energy photons produced by the optical region electrons as
presented in section \ref{sec:gamma-opt} (i.e., $N_{\rm
det,\gamma-opt}=0$) if $R_{\rm \gamma}>R_{\rm opt}$.

\subsection{Very early EIC emission}\label{sec:EIC}
Whatever the mechanism is, the prompt emission should have an
internal origin, in view of the high variability of the light curves
and the very sharp decline at $t>T_{90}$. External reverse-forward
shock formed very quickly. Consequently, the prompt photons passing
through the reverse/forward shock regions were IC scattered by the
shock accelerated electrons. As a result, two additional GeV-TeV EIC
components were present.

\subsubsection{EIC in reverse shock region}
 \citet{Racusin08} and \citet{Wu08} argued that the reverse shock
 emission of the narrow jet component had not been seen. Its physical
 parameters are thus unknown. {In some optical flash modeling, the
$\epsilon_{\rm B}$ (or/and $\epsilon_{\rm e}$) of RS is found to be
much larger than that of the FS (Fan et al. 2002; Zhang, Kobayashi
\& M\'esz\'aros 2003; Kumar \& Panaitescu 2003; cf. Nakar \& Piran
2005). However, if such a phenomena is popular very bright optical
flashes would be frequently detected (McMahon, Kumar \& Piran 2006),
inconsistent with current optical afterglow observations. For the
particular burst GRB 080913B, Racusin et al. (2008) argued that the
RS of the wide jet component has an $\epsilon_{\rm B} \sim 0.1$,
much larger than that of the corresponding FS. However, if
$\epsilon_{\rm B} \sim 0.1$ holds for the RS of the narrow core too,
the resulting optical emission would be too strong to match the data
(X. F. Wu. 2008, private communication). On the other hand, assuming
that these two parameters are the same as those of the forward
shock, it is straightforward to show that the RS optical emission of
the narrow core is $\sim 0.3$ Jy at the crossing time, outshone by
the simultaneous prompt emission and consistent with the data. So
below we simply assume that the shock parameters of the FS and RS
are the same for the narrow jet component.}

The reverse shock emission must have
 overlapped the prompt gamma-rays and optical emission. Therefore
 the electrons accelerated by the reverse shock front were
 cooled by the prompt emission and gave rise to an EIC radiation
 component \citep{b05,fzw05}.

The number of electrons in the reverse shock region is
\begin{equation}
N_{\rm e,r} \simeq \frac{E_{\rm k,n}}{\Gamma_{\rm } m_p c^2} \simeq
3 \times 10^{55} E_{\rm k,n,55.5} \Gamma_{\rm 2.8}^{-1}.
\end{equation}

The typical radius of the reverse shock can be estimated as
\begin{equation}
R_{\rm r} \sim 2\Gamma^2 c T_{90}/(1+z) \sim 5\times 10^{17}
\Gamma_{2.8}^2 ~{\rm cm}.
\end{equation}

The optical depth of the prompt photons being scattered by the electrons was
\begin{equation}
\tau_{\rm r} \sim \sigma_{_{\rm T}}{N_{\rm e,r}\over 4 \pi R_{\rm
r}^2} \sim 7 \times 10^{-6}E_{\rm k,n,55.5} \Gamma_{\rm
2.8}^{-1}R_{17.7}^{-2}.
\end{equation}

On the other hand, the total number of the prompt soft $\gamma$-rays
that reached us (per area) can be estimated as \citep{fp06}
\begin{equation}
N_{\rm tot,\gamma} \sim  {\beta_{_{\gamma}}-1 \over
\beta_{_{\gamma}}}{{\cal F} \over h\nu_{\rm \gamma,p}},
\label{eq:N_tot}
\end{equation}
where ${\cal F} \sim 10^{-4}~{\rm erg~cm^{-2}}$ is the energy
fluence of the prompt $\gamma$-rays and $\beta_\gamma \sim 2.6$ is
the high energy spectral index of the prompt $\gamma$-ray emission.

The number of the reverse shock EIC photons detectable by the Fermi satellite
and their typical energy can be estimated as
\begin{equation}
N_{\rm det,r} \sim \tau_{\rm r} N_{\rm tot,\gamma} S_{\rm det} \sim
3,
\end{equation}
and
\begin{equation}
h\nu_{_{\rm EIC},\rm r} \sim 2 \gamma_{\rm m,r}^2 E_{\rm p} \sim
13~{\rm GeV}~({\gamma_{\rm m,r} \over 100})^2,
\end{equation}
where $\gamma_{\rm m,r}$ \footnote{$\gamma_{\rm m,r} \sim
\epsilon_{\rm e,n} (\bar\gamma-1) m_{\rm p}/m_{\rm e} (p_{\rm
n}-2)/(p_{\rm n}-1)$, where $m_{\rm p}$ is the rest mass of the
protons, and $\bar\gamma$ indicates the internal energy density in
the shocked region.  For the mildly Relativistic reverse shock we
have $\bar\gamma-1 \sim 1$ \citep[see][for details]{zwd05}, which is
the case for this burst.
}
 is the minimal Lorentz factor of the electrons
accelerated in the reverse shock front. The electrons are in slow
cooling phase since the cooling Lorentz factor is \citep{fp08}
\begin{equation}
\gamma_{\rm c,r} \sim 10^{3} \Gamma_{2.8}^3 R_{\rm
r,17.7}L_{\gamma,52.7}^{-1}>\gamma_{\rm m,r}.
\end{equation}
The Compton parameter $Y_{\rm EIC,r} \sim \gamma_{\rm m,r}^2
\tau_{\rm r} \ll 1$. So the energy of this EIC component was much
smaller than that of the prompt soft $\gamma$-rays.

Here we do not take into account the cooling caused by the
synchrotron radiation because $U_B \sim  \varepsilon_B 4 \Gamma^2 A
R^{-2} m_p c^2 \sim 3.3\times 10^{-3}  \varepsilon_{\rm n,B,-4}
\Gamma_{2.8}^2 A_{\star,-2} R_{17.7}^{-2}{\rm erg}\,{\rm cm}^{-3}$,
which is much smaller than $U_{\gamma} \sim \frac{L_{\gamma}}{4\pi
R^2 \Gamma^2 c} \sim 1.3 \Gamma_{2.8}^{-2} R_{17.7}^{-2}{\rm
erg}\,{\rm cm}^{-3}$.

Some prompt optical photons will be up-scattered  by the reverse
shock electrons and will be boosted to an energy $\sim
2\gamma_{m,r}^2 h\nu_{\rm opt} \sim 10$ keV, which is too low to be
of interest.

\subsubsection{EIC in forward shock region}
The prompt emission will
cool the forward shock electrons as well \citep{fzw05,wp06}.
However, for the prompt $\gamma$-rays, this EIC process is
unimportant since it is in the Klein-Nishina regime. Because
the large radius lowers the optical depth, pair
avalanche does not exist in this case. Here we focus
on the EIC radiation of the prompt optical emission. The
energy density of the emitted prompt photons is $U_{\rm opt} \sim
\frac{L_{\rm opt}}{4 \pi R^2 \Gamma^2 c} \sim 0.05~L_{\rm
opt,51}\Gamma_{2.8}^{-2} R_{17.7}^{-2}{\rm erg}\,{\rm cm}^{-3}$,
which is larger than $U_{\rm B}$. So the cooling of the forward
shock electrons is dominated by the EIC process.

The number of the electrons swept by the forward shock is
\begin{equation}
N_{\rm e,f} \simeq 4 \pi A R \simeq 1.8 \times 10^{52} A_{\star,-2}
R_{17.7}.
\end{equation}
The optical depth of the prompt photons for being scattered is thus
\begin{equation}
\tau_{\rm f} \sim \sigma_{_{\rm T}}{N_{\rm e,f}\over 4 \pi R^2} \sim
3\times 10^{-9}A_{\star,-2} R_{17.7}^{-1}.
\end{equation}

Noticing that we don't know the spectrum in the optical band, {we
can only evalute the lower limit by taking the observed optical
emission as the peak.} The total number of the optical photons
reaching us (in unit area) can be estimated as
\begin{equation}
N_{\rm tot, opt} \sim  {{\cal F}_{\rm opt} \over h\nu_{\rm opt}}\sim
10^6 {\rm cm}^{-2}, \label{eq:N_tot_opt}
\end{equation}
where ${\cal F}_{\rm opt}$ is the fluence of the prompt optical
emission.

\begin{figure}
\includegraphics[width=0.5\textwidth]{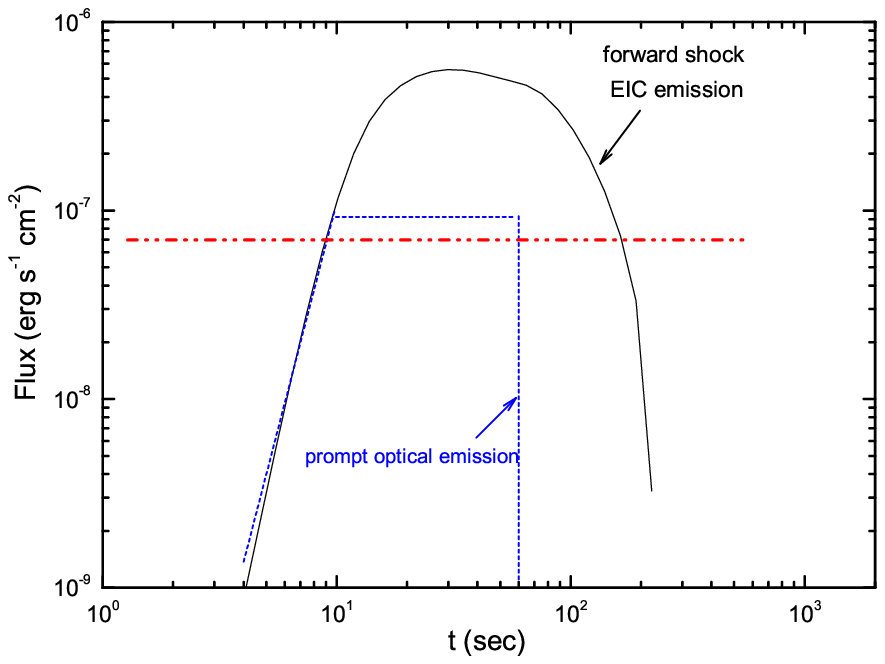}
\caption{The numerical light curve of GRB 080319B forward shock EIC
emission, and the prompt optical prototype is also shown. {The dash
dot-doted line represents the detection of five $\sim 10$ GeV
photons in $120$ sec by LAT onboard Fermi satellite.}}
\label{fig:eic}
\end{figure}

For Fermi, the detectable number of the forward shock EIC radiation
can be estimated as
\begin{equation}
N_{\rm det,f} \sim \tau_{\rm f} N_{\rm tot,opt} S_{\rm det} \sim 30.
\end{equation}
{Usually for an integration time $t_{\rm int}\lesssim 10^{5}$ sec,
LAT needs 5 high energy photons to claim a significant detection
\citep[e.g.,][]{zm01,fzw05}. With a duration of 120 s, and for the
typical energy of $\sim 10$ GeV, this detection corresponds to $8
\times 10^{-8} {erg/s/cm^2}$. We plot such a threshold in
Fig.\ref{fig:eic} and find out that the forward shock EIC emission
component is detectable in $\sim 100$ s, longer than the prompt soft
gamma-ray emission.}

The typical energy of these forward shock EIC photons is
\begin{equation}
h\nu_{_{\rm EIC},\rm f} \sim 2 \min\{\gamma_{\rm c,f}^2,\gamma_{\rm
m,f}^2\} h\nu_{\rm opt} \sim 10~{\rm GeV},
\end{equation}
where $\gamma_{\rm c,f}\sim 10^5 \Gamma_{2.8}^3R_{17.7}L_{\rm
opt,51}^{-1}$ and $\gamma_{\rm m,f} \sim 4\times 10^4 \Gamma_{2.8}$.

The Compton parameter $Y_{\rm EIC,f} \sim \gamma_{\rm m,f}^2
\tau_{\rm f} \sim 10$. As the emitted energy of the optical photons was
$3\times 10^{52}$ergs (isotropic), the total energy of the EIC
photons by forward shocked electrons is $\sim 3\times 10^{53}$ergs.

In the rest frame of the forward shock, the seed optical photons
have a typical energy $\sim \gamma_e h\nu_{\rm opt}/\Gamma < m_e
c^2$ for $\gamma_e<10^{8} \Gamma_{2.8}(h\nu_{\rm opt}/2{\rm
eV})^{-1}$. So the EIC scattering in the forward shock front is well
in the Thompson regime. The resulting spectrum for $\nu<\nu_{_{\rm
EIC},\rm f}<1$ TeV is expected to be not steeper than  $F_\nu
\propto \nu^{-p_{\rm n}/2} \sim \nu^{-1.2}$. On the other hand, the
absorption depth for a 30 GeV photons from a redshift $z\sim 1$ is
only about 1 \citep{stecker06}. So we expect that, if Fermi worked
at that moment, it could have detected some photons as energetic as
$\sim 30$ GeV.

Though very bright optical flashes from GRBs are very rare, a few
such events are still possible during Fermi's 10 years of operation.
Since the EIC component from the forward shock
region can give rise to a significant detection for a Fermi-like
satellite, here we use the numerical code by \citet{fan08} to
a more detailed estimate. For simplicity, we approximate the prompt optical
emission flux by $F=10^{-7}(t/10)^6~{\rm erg~s^{-1}~cm^{-2}}$ for
$t<10$ sec, a constant plateau lasting till $t\sim 60~{\rm sec}$,
and $F=0$ afterward. The optical spectrum is set as a typical Band
function \citep{band93}, for which (the break energy, the low energy
spectral index, the high energy spectral index) are taken as (2 eV,
-1, -2.25), respectively. Notice that it is also a lower limit, as
we take the V band as the peak. As shown in Fig. \ref{fig:eic}, the
forward shock EIC emission lasts about twice that of the
prompt emission. This is because the duration of the high-energy
emission is affected by the spherical curvature of the blast wave
\citep{b05} and is further extended by the highly anisotropic
radiation of the up-scattered photons \citep{fp06,wp06}. We also
find out that the total energy of the EIC emission is about 10 times
that of the prompt optical emission, consistent with our analytical
estimate.

\subsection{The late GeV-TeV emission of the external forward shock}
The high energy emission of the external forward shock has been
extensively discussed in literature since 1994 (M\'esz\'aros \& Rees
1994; Dermer, Chiang \& Mitman 2000; Sari \& Esin 2001; Wang, Dai \&
Lu 2001; Zhang \& M\'esz\'aros 2001; Fan et al. 2008).
GRB 080319B is distinguished from most bursts
by its huge $E_{\rm k,n}$ and by the
large contrast between $\epsilon_{\rm n, e}$ and $\epsilon_{\rm
n,B}$, both indicating a very strong high energy radiation
component.

In the very early afterglow phase ($t\leq 60$ s), the Lorentz factor
of the forward shock is almost a constant. The typical Lorentz factor
of the shocked electrons is $\gamma_{\rm m} \sim 4\times 10^4
\epsilon_{\rm e,n,-1} \Gamma_{2.8}$.

After that, the forward shock forms a self-similar profile and its
Lorentz factor can be estimated as
\begin{eqnarray}
\Gamma \sim 310 (1+z)^{1/4} E_{\rm k,n,55.5}^{1/4}
A_{\star,-2}^{-1/4} t_3^{-1/4}.
\end{eqnarray}
The typical Lorentz factor of the shocked electrons is
\begin{eqnarray}
\gamma_{\rm m} \sim 2 \times 10^4 (1+z)^{1/4} \epsilon_{\rm e,n,-1}
E_{\rm k,n,55.5}^{1/4} A_{\star,-2}^{-1/4} t_3^{-1/4}.
\end{eqnarray}

At this stage, the forward shock is in the slow cooling phase
\citep{Racusin08}, and $\nu_{\rm m} < \nu_{\rm X} <\nu_{\rm BAT}
<\nu_{\rm c}$, where $\nu_{\rm BAT} \sim 10^{20}$ Hz is the
frequency of the BAT detector onboard {\it Swift} satellite and
$\nu_c$ is the cooling frequency \footnote{For $\nu>\nu_{\rm c}$,
the synchrotron radiation spectrum is $\propto \nu^{-p_{\rm n}/2}$.
On the other hand, the maximum synchrotron radiation frequency
$h\nu_{\rm M} \sim 30\Gamma/(1+z)~{\rm MeV}$ \citep{cw96} is up to a
few GeV for $\Gamma \geq 300$. It is straightforward to show that
the fluence of the high energy afterglow emission in the energy
range of LAT is in order of $10^{-5}~{\rm erg~cm^{-2}}$, comparable
to the fluence of the soft X-ray/$\gamma$-ray afterglow emission.
Such a conclusion is almost independent of the afterglow models.}.
On the other hand, $\Gamma m_e c^2/\gamma_m \sim m_e c^2/100 \sim
10^{18}$Hz $< \nu_c$, implying that the SSC emission of the
electrons with a Lorentz factor $\sim \gamma_c$ is in extreme
Klein-Nishina regime and it is effectively suppressed. So we expect
that the SSC emission will peak at an energy
\begin{eqnarray}
h\nu_{\rm p}^{\rm SSC} &\sim &  \Gamma \gamma_m m_e c^2
\nonumber \\
& \sim & 2.3 ({1+z \over 2})^{1\over 2} \epsilon_{e,-1} E_{\rm
k,n,55.5}^{1/2} A_{\star,-2}^{-1/2} t_3^{-1/2} {\rm TeV},
\end{eqnarray}
for {the late afterglow.} The SSC emission of the forward shock
in the very early afterglow phase overlap with the GeV-TeV emission
of the {prompt phase} and is very likely to be outshone. Below we
just discuss the SSC emission of the forward shock in the normal
decline phase ($t>60$ sec).

\begin{figure}
\includegraphics[width=0.5\textwidth]{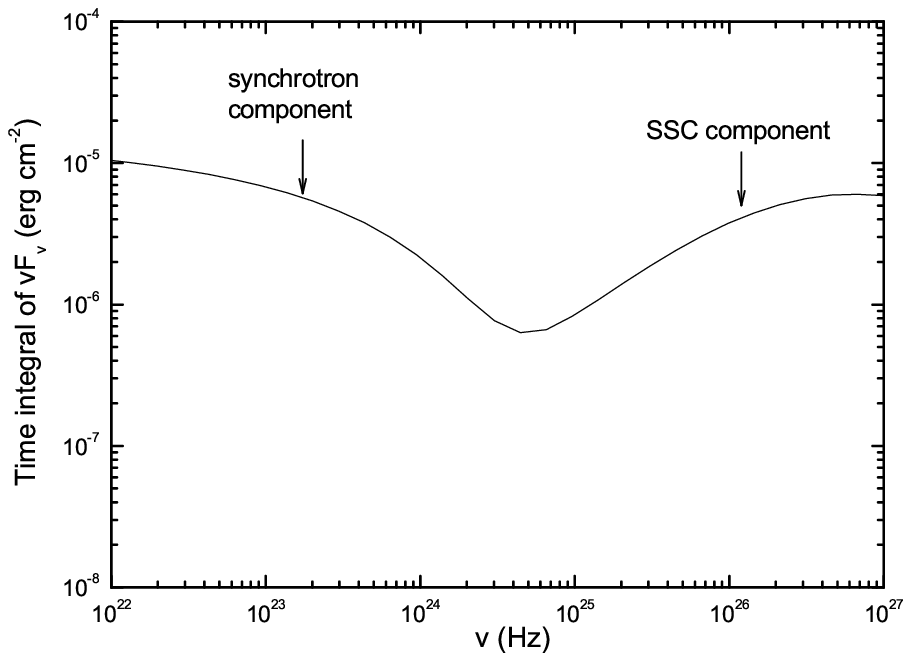}
\caption{The integral of $\nu F_\nu$, including the synchrotron +
SSC components, in the time interval 60 sec $-$ 2000 sec.}
\label{fig:essc}
\end{figure}

To check our estimate, we calculate numerically with Fan et al.'s code (2008)
the forward shock emission spectrum. As shown in
Fig.\ref{fig:essc}, the SSC emission peaks at TeV energies, with
a fluence $\sim 6 \times 10^{-6} {\rm erg\, cm^{-2}}$, and
an isotropic energy $\sim 3 \times 10^{52} {\rm erg}$.
The detection of the TeV emission is beyond the scope of the Fermi
satellite. Ground based Cherenkov telescopes, like MAGIC and
H.E.S.S, may be suitable to detect these energetic signals. However,
before reaching us, these TeV photons would have been absorbed by
the infrared background photons, and such emission could be seen
only from rare very nearby sources.

We find in Fig.\ref{fig:essc} that for a Fermi-like satellite the
MeV-GeV synchrotron radiation of the forward shock may give rise to
a detectable signal. In our calculation, we take a maximal Lorentz
factor of the shocked electrons $\gamma_{\rm M} \sim 4\times 10^7
B^{-1/2}$ \citep{cw96}, where $B$ is the magnetic field generated in
the shock front. This leads to the synchrotron GeV cutoff (see
Fig.\ref{fig:essc}). As a numerical example, following
\citet{fan08}, we take a real effective area of LAT and integrate
the spectrum over the frequencies to estimate the number of
detectable photons. For this particular example, the LAT onboard
Fermi can detect $\sim 400$ ( $>20$ MeV), $\sim 20$ ( $>1$ GeV), and
$\sim 0.1$ ( $>100$ GeV, without the correction due to the
absorption by the infrared background photons) high energy photons.
{This would be a very exciting detection.}

\section{Origins of GeV emission of some recent GRBs}
Recently high energy emission has been detected by AGILE: GRB 080514B
\citep{Giuliani08}, and by Fermi: GRB 080825C \citep{Bouvier08}, GRB 080916C
\citep{Tajima08} and GRB 081024B \citep{Omodei08}.
 We can apply the above considerations for GRB 080319B
to all these bursts, though the very early afterglow data
are unavailable and the constraints on the model are not very tight.

\emph{GRB 080514B}: the burst light curve shows a multi-peaked
structure with a duration of $\sim 7$s \citep{Golenetskii08}. The high energy
emission lasted about 2 times longer than the MeV emission
 and the most intense high energy emission arrived
at $\sim 10$ sec after the trigger \citep{Giuliani08}. We interpret
such an intense high energy flash as the EIC emission in the reverse
shock region. In this case, some seed photons (the prompt MeV
emission) are upscattered by the reverse shock electrons and are
boosted to an energy $\lesssim 1$ GeV \citep{b05,fzw05}. The main
advantage of this model is that the duration of the high energy
emission is longer than that of the prompt soft $\gamma$-ray
emission by a factor of 2, consistent with the observation.
There are also 2 high energy photons detected at $\sim 26$ s.
They may be the synchrotron or SSC emission of the forward shock.
The possibility that they are the SSC emission of an underlying
X-ray flare (Wei, Yan \& Fan 2006; Wang, Li \& M\'esz\'aros 2006;
Galli \& Guetta 2008; Fan et al. 2008) cannot be ruled out. The
lack of the simultaneous XRT observation makes it difficult to draw further
conclusion.\\

\emph{GRB 080916C} was a long burst with a duration $\sim 60$s. The
time averaged spectrum, from 8 keV up to 30 MeV, of the main
emission is best fitted by a Band function with $E_{\rm p} = 424\pm
24$keV, $\alpha = -0.91\pm 0.02$, and $\beta = -2.08 \pm 0.06$. The
fluence (8 keV $-$ 30 MeV) is $1.9\times 10^{-4} {\rm erg/cm^2}$
\citep[Swift]{van08} \citep[slightly different in Konus-Wind
observation, ][] {Golenetskii08}. More than 10 photons are observed
above 1 GeV during the prompt phase
 \citep{Tajima08}
and the high energy emission lasted longer than the soft $\gamma-$rays
 \citep{Abdo09}.
This was a very bright burst with a hard spectrum. A simple extension
of the keV$-$MeV spectrum to higher energy range gives $N({\rm 30
MeV-1 GeV})\sim 700$, $N({\rm 1-10 GeV})\sim 100$ and $N({\rm >10
GeV})\sim 9$ by LAT (on-axis case), enough to match the observation
\citep{Tajima08,Omodei08b}. This fact suggests that the synchrotron
radiation of the internal shocks plays an important role in
producing high energy prompt emission.

{The redshift of GRB 080916C is estimated to be $\sim 4.5\pm 0.1$
\citep{Greiner09}. The peak luminosity is thus as high as $\sim
5\times 10^{53}~{\rm erg~s^{-1}}$. The typical variability timescale
is suggested to be as long as $\sim 2$ s \citep{Abdo09}. With these
information,} the detection of $>10$ GeV prompt emission gives a
tight constraint on the initial bulk Lorentz factor of the GRB
outflow, i.e. (Lithiwick \& Sari 2001; Fan \& Piran 2008; Gupta \&
Zhang 2008),
\[\Gamma_{\rm i}>400 ({h\nu_{\rm
cut}\over 10{\rm GeV}})^{p\over 2(p+4)}L_{\gamma,54}^{1\over
p+4}\delta t_{\rm 0}^{-{1\over p+4}}.
\]
Using the maximal synchrotron radiation frequency of the shocks
$h\nu_{\rm M}\approx 30\Gamma/(1 + z)~{\rm MeV}$ \citep{cw96}, we
find that if the high energy emission up to $\sim 10$ GeV is
attributed to the synchrotron radiation of internal shocks, the
initial Lorentz factor should satisfy:
\[ \Gamma_{\rm i}\geq 1800
{1+z\over 5.5}{h\nu_{\rm cut} \over 10~{\rm GeV}}.
\]
A $\Gamma_{\rm i}$ much higher than $2000$ is unlikely. So this
strongly suggests that the internal shocks can accelerate high
energy particles (both protons and electrons) very efficiently and
the highest energy of electrons is limited by the loss via
synchrotron radiation. The energy distribution index of the
shock-accelerated electrons ($p\sim 2.4$) is also close to that
predicted in the theory. This is a very encouraging news for the
people interested in the ultra-high energy particle acceleration in
GRBs. However, we'd like to caution that it is the only case among
the 70 events observed so far by Fermi \citep{Abdo09}. It might be
too early to say more at this moment.

The internal shock synchrotron radiation cannot account for the
delayed high energy emission \citep{Abdo09}. The possible mechanisms
that can produce this emission are (i) the EIC emission from the
reverse-forward shock regions, (ii) the SSC emission of the forward
shock and (iii) SSC emission of the weak internal shocks powering an
extended X-ray emission component that is below the threshold of
GBM.

{The spectrum of the $\geq 100$ MeV emission in the time interval
$\sim 200-1400$ sec is $F_\nu \propto \nu^{-1.8\pm 0.5}$
\citep{Abdo09}. Such a soft spectrum imposes a tight constraint on
the models. In the standard afterglow model, the late time infrared
and X-ray afterglow \citep{Greiner09} can only be interpreted as the
forward shock emission of an ejecta expanding into a weak stellar
wind. Like in GRB 080319B, an $A_* \sim 0.01$ is needed to have a
cooling frequency above the XRT energy range at $t\geq 1$ day (Gao
et al. 2009, in preparation). An electron energy distribution index
$p \sim 2.2$ can reproduce both the infrared to X-ray spectrum
$F_\nu \propto \nu^{-0.63}$ and the X-ray (infrared) afterglow
decline $\propto t^{-1.29\pm 0.09}$ ($t^{-1.40\pm 0.05}$). The
spectrum of the SSC or the EIC emission\footnote{For the EIC model,
the flare photons should peak in far ultraviolet band, as suggested
in \citet{fp06}, otherwise the typical energy of the scattered
photons will be well above $\sim 100(1+z)$ MeV and we need a very
strong X-ray flare to account for the detected high energy photons.}
of the forward shock should have a spectrum not steeper than
$\nu^{-p/2} \sim \nu^{-1.1}$, and can only marginally match the
data. So we prefer the possibility (iii). For the X-ray emission
powered by the prolonged activity of the central engine, the SSC
emission can peak at an energy $\lesssim 550[(1+z)/5.5]$ MeV
\citep[see section 5.1 of][for details]{fan08}. In this case, the
electron energy distribution index is irrelevant to that of the
afterglow electrons and can be as large as $\sim 3$, as found in the
spectrum analysis of X-ray flares \citep{Butl07}. As a result, the
soft spectrum of the delayed $>100$ MeV
emission may be interpreted.}\\

\emph{GRB 081024B} was a short burst with a duration $\sim 0.4-0.8$s
\citep{Connaughton08,Hanabata08}. The LAT saw the emission from this
source up to 3 GeV, in the first 5 seconds after the trigger.
Here we consider two possible interpretations. One is that the
delayed emission is the SSC component of an extended/prompt soft
X-ray emission. Following Fan \& Piran (2008; see their
eqs.(47-49)), the typical frequency of the internal shock SSC
emission can be estimated as
\[
h\nu_{\rm m}^{\rm ssc} \sim 75~{\rm MeV}~(E_{\rm p}/0.3~{\rm
keV})^2R_{\rm int,14}L_{\rm X,49}^{-1/2}(1+Y_{\rm ssc})^{1/2},
\]
where $R_{\rm int}$ is the radius of the continued but weak internal
shocks that power the underlying prompt X-ray emission with a
luminosity $L_{\rm X}$, and $Y_{\rm ssc}$ is the SSC parameter of
the internal shocks. This model requires a unmagnetized outflow
launched by the continued activity of the central engine, in
contradiction with most models proposed so far (see Zhang 2006 for a
review). If confirmed, a stringent constraint on the nature of the
extended emission following short GRBs will be established. So, in
principle, the cooperation of {\it Swift} and Fermi satellite can
reveal the nature of the late outflow powering the extended
emission. The other possible origin of the delayed high energy
emission is the SSC emission of the forward shock. It is
straightforward to show that the outflow with an initial Lorentz
factor $\Gamma_{\rm i} \sim 400$ gets decelerated in the
interstellar medium with a number density $\sim 1~{\rm cm^{-3}}$ in
$\sim 5$ sec. The typical SSC emission frequency of the forward
shock can be estimated as \citep{se01,fp08}
\[
h\nu_{\rm m}^{\rm ssc} \sim 25 ~{\rm GeV}~\epsilon_{e,-1}^4
\epsilon_{\rm B,-2}^{1/2}\left[{13(p-2)\over 3(p-1)}\right]^4E_{\rm
k,51}^{3/4}t_{1}^{-9/4}.
\]
One may be able to distinguish between the above two scenarios by
analyzing the spectrum.
If the delayed high energy emission is the
SSC component of extended but weak internal shocks, the $0.1-3$ GeV
spectrum is expected to be steeper than $\nu^{-1}$. If the delayed
high energy emission is the SSC component of external forward shock,
the $0.1-3$ GeV spectrum is expected to be
$\nu^{-1/2}$ unless $p\sim 2$. The forward shock synchrotron
radiation can also give rise to GeV emission. It is, however, difficult to
say more concerning this possibility because the early afterglow
physics of short GRBs is still poorly understood.

\section{Conclusions and discussions}

\begin{table}
\caption{The expected emission of high energy photons {(mostly
inverse Compton scattering)} from different origins for GRB 080319B,
which should be detectable by LAT onboard Fermi satellite.}
\begin{tabular}{l|l|c|c|c|c}
\hline
Seeds & region for& duration (s) & typical photon  & detectable\\
 & electrons  & & energy (GeV) &  photons
\\ \hline
$\gamma$-rays & prompt opt & $\sim$ 60 &  $\gtrsim 8$ & $\leq 100$
\\ \hline
opt & prompt $\gamma$-ray & $\sim$ 60 &  $\sim$ 15 & $\sim 240$ $^\dagger$
\\ \hline
$\gamma$-rays &  reverse shock & $\sim 10^2$ & $\sim$ 13 & $\sim$ 3
\\ \hline
opt &  forward shock & $\sim 10^2$ & $\sim$ 10 & $\sim$ 30
\\ \hline
afterglow & {(Synchrotron)} & $\sim 10^3$ & 0.01-0.1 &  $\sim 400$
\\ \hline
afterglow & external shock & $\sim 10^3$ & $\sim 10^3$ & $\sim$ 0.04 $^\ddagger$
\\ \hline
\multicolumn{5}{p{0.45\textwidth}}{\footnotesize $^\dagger${This
case is less likely, and possibly alternates with the former case.}}
\\
\multicolumn{5}{p{0.45\textwidth}}{\footnotesize
$^\ddagger${Supposing an instrument with effect area $10^4 {\rm
cm^{2}}$ and without considering the absorption on the way to the observer.}}
\\ \hline
\end{tabular}
\label{tab:con}
\end{table}

High-energy emission provides a new window into prompt
emission/afterglow physics and can provide an independent test of
models. Motivated by this, we calculate the possible high-energy
prompt/afterglow emission in GRB 080319B that was distinguished by a
naked-eye optical flash and by an unusual strong early X-ray
afterglow. Two possible GeV-TeV emission components may be related
to the naked-eye optical flash. The first is the Inverse Compton
scattering of the prompt optical photons by electrons producing the
soft $\gamma$-rays. The second is the very early EIC emission from
the forward shock region when the prompt optical emission overlaps
the shock front. The difference is their duration. The former is
expected to be simultaneous with the prompt soft $\gamma$-ray
emission while the latter lasts longer (see Fig.\ref{fig:eic}). The
synchrotron radiation of the forward shock can give rise to a
significant detection, too (see Tab. \ref{tab:con} for a summary).
This component may be more common than the two that depend on a
strong optical flash as which is quite rare. The detection prospect
of the forward shock synchrotron radiation by LAT is fairly good.
For the {\it Swift} GRBs detected so far, GRBs 060105, 061007,
070419B and 080721 have a $0.3-10$ keV flux $\sim 10^{-8}~{\rm
erg~s^{-1}~cm^{-2}}$ at $t\sim 100$ s after the trigger
(http://www.swift.ac.uk/xrt$_{-}$curves/; Evans et al. 2007). Though
about one order of magnitude lower than that of GRB 080319B, they
are strong enough to produce a GeV synchrotron emission detectable
by LAT as long as the synchrotron spectrum can indeed extend to an
energy $\sim 30\Gamma/(1+z)~{\rm MeV}$. The forward shock SSC
emission of these very bright events may be {more} suitable for the
ground-based Cherenkov telescopes, like MAGIC or H.E.S.S.

In section 3, we discussed the {possible} physical origin of the
high energy emission of GRB 080514B, GRB 080916C and GRB 081024B. We
find that these detections can be {generally} understood by the
synchrotron and inverse Compton radiation of the internal shocks or
external shocks. For example, the delayed sub-GeV flash detected in
GRB 080514B may be the EIC emission from the reverse shock region
and the prompt GeV-emission of GRB 080916C may be dominated by the
synchrotron radiation of the internal shocks. The ``long lasting"
high energy emission detection in the short burst GRB 081024B may be
attributed to the SSC emission of the decelerated forward shock or
the internal shocks powering an extended X-ray component which is
below the threshold of GBM. {However, as lack of detailed
observations, it is defficult to draw a firm conclusion.}

Finally we focus on the common feature that the high energy emission
usually lasts longer than the prompt soft $\gamma$-rays, as detected
in GRB 080514B, GRB 080916C and GRB 081024B. Such a phenomena,
peculiar in pre-afterglow era, {may be explained as}: (1) The
synchrotron and the SSC emission of the long lasting forward
external shock can contribute to the high energy emission
significantly. (2) The GRB central engines usually do not turn off
abruptly. The SSC emission of the continued but weak internal shocks
may peak at GeV energies. (3) If a (mildly) relativistic reverse
shock formed, the prompt optical/X-ray/$\gamma$-ray photons overlap
the external shock fronts tightly and cool the accelerated electrons
effectively. This process will produce a GeV emission component with
a duration about twice that of the prompt photons. For a
sub-relativistic reverse shock, the prompt soft $\gamma$-ray photons
exceed the external shock fronts quickly. Its effect on cooling the
reverse/forward shock electrons can be ignored.
 However in such a case the
electrons/protons accelerated in reverse shock contain just
$\lesssim 10\%$ of the total energy of the GRB ejecta
\citep{NP04,Mimica08} and cannot play an important role in producing
high energy emission. (4) The EIC in the late afterglow phase caused
by X-ray flares can also give rise to GeV emission. However the
luminosity is lowered since its duration has been significantly
extended. Usually LAT is unable to catch such a weak signal.

\section*{Acknowledgments}
YZF thanks Bing Zhang and Daming Wei for discussions.
This work is supported in part by the Israel Science Foundation (for
TP), a grant from the Danish National Science Foundation, a special
grant of Chinese Academy of Sciences, National basic research
program of China grant 2009CB824800, and the National Natural
Science Foundation of China under the grants 10673034 (for YZF) and
10703002 (for YCZ).


\begin{thebibliography}{}
\bibitem[Abdo et al.(2009)]{Abdo09} Abdo A. A., et al., 2009, Science, in press
\bibitem[Band et al.(1993)]{band93} Band D. et al., 1993, ApJ, 413, 281
\bibitem[Beloborodov(2005)]{b05}Beloborodov A. M., 2005, ApJ, 618, L13
\bibitem[Blandford \& McKee (1976)]{BM76} Blandford R. D., McKee C. F., 1976, Phys. Fluids., 19, 1130
\bibitem[Bloom et~al.(2008)]{Bloom08} Bloom J. S., et~al., 2008, ApJ submitted
(arXiv:0803.3215)
\bibitem[Bouvier et~al.(2008)]{Bouvier08}Bouvier A., et~al., 2008, GCN circular, 8183
\bibitem[Butler \& Kocevski (2007)]{Butl07} Butler N. R., Kocevski D.,
2007, ApJ, 663, 407
\bibitem[Cheng \& Wei(1996)]{cw96} Cheng K. S., Wei D. M., 1996,
MNRAS, 283, L133
\bibitem[Chevalier \& Li (2000)]{CL00} Chevalier R. A., Li Z. Y., 2000, ApJ, 536, 195
\bibitem[Connaughton et~al.(2008)]{Connaughton08}Connaughton V., et~al., 2008, GCN circular, 8408
\bibitem[Cwiok et~al.(2008)]{Cwiok08} Cwiok M., et al., 2008, GCN circular, 7445
\bibitem[Dai \& Lu (1998)]{DL98} Dai Z. G., Lu T., 1998, MNRAS, 298, 87
\bibitem[Dermer et al. (2000)]{dcm00}Dermer C. D., Chiang J., Mitman K. E., 2000, ApJ, 537, 785
\bibitem[Evans et al. (2007)]{Evans07} Evans P. A., et al., 2007, A\&A, 469, 379
\bibitem[Fan et al. (2002)]{fan02} Fan Y. Z., Dai Z. G., Huang Y.
F., Lu T., 2002, ChJAA, 2, 449
\bibitem[Fan \& Piran (2006)]{fp06}Fan Y. Z., Piran T., 2006,
MNRAS, 370, L24
\bibitem[Fan \& Piran (2008)]{fp08}Fan Y. Z., Piran T., 2008,
Front. Phys. Chin., 3, 306 (arXiv:0805.2221)
\bibitem[Fan et al.(2008)]{fan08}Fan Y. Z., Piran T., Narayan R., Wei D. M., 2008,
MNRAS, 384, 1483
\bibitem[Fan, Zhang \& Wei(2005)]{fzw05}Fan Y. Z., Zhang B., Wei D. M., 2005, ApJ, 629,
334
\bibitem[Fan, Zhang \& Wei(2009)]{fzw09}Fan Y. Z., Zhang B., Wei D. M., 2009, Phys. Rev. D.,
79, 021301
\bibitem[Galli \& Guetta (2008)]{gg08}Galli A., Guetta D. A., 2008, A\&A, 480, 5
\bibitem[Giuliani et al.(2008)]{Giuliani08}Giuliani A., et al., 2008,
A\&A, 491, L25
\bibitem[Golenetskii et al.(2008)]{ga08}Golenetskii, S., Aptekar, R., Mazets, E., Pal'shin, B., Frederiks, D., Cline, T., 2008, GCN circualr, 7482
\bibitem[Golenetskii et al.(2008a)]{Golenetskii08}Golenetskii S., et~al., 2008, GCN circular 7751
\bibitem[Golenetskii et al.(2008b)]{Golenetskii08b}Golenetskii S., et~al., 2008b, GCN circular 8258
\bibitem[Gonz\'alez et al. (2003)]{Gonz03}Gonz\'alez M. M., Dingus B. L., Kaneko Y., Preeze R. D., Dermer C. D., Briggs M. S., 2003, Nature, 424, 749
\bibitem[Greiner et al. (2009)]{Greiner09} Greiner J., et al., 2009,
A\&A (arXiv:0902.0761)
\bibitem[Gupta \& Zhang (2008)]{GZ08} Gupta N., Zhang B., 2008,
MNRAS, 384, L11
\bibitem[Hanabata et~al.(2008)]{Hanabata08}Hanabata Y., et~al., GCN circular, 8444
\bibitem[Hurley et~al.(1994)]{Hurley94}Hurley K., Dingus B. L., Mukherjee R., et~al., 1994,  Nature, 372, 652
\bibitem[Jin \& Fan (2007)]{JF07} Jin Z. P., Fan Y. Z., 2007, MNRAS, 378,
1043
\bibitem[Jin et~al.(2007)]{Jin07} Jin Z. P., Yan T., Fan Y. Z., Wei
D. M. 2007, ApJ, 656, L57
\bibitem[Karpov et~al.(2008)]{Karpov08}Karpov S., et~al., 2008, GCN Circular 7502
\bibitem[Kumar \& Panaitescu (2003)]{KP03} Kumar P., Panaitescu A., 2003,
MNRAS, 346, 905
\bibitem[Kumar \& Panaitescu(2008)]{kp08}Kumar P.,  Panaitescu, A., 2008, MNRAS, 391, L19
\bibitem[McMahon et~al.(2006)]{McMa06} McMahon E., Kumar P., Piran
T., 2006, MNRAS, 366, 575
\bibitem[Margutti et~al.(2008)]{Margutti08} Margutti R., Guidorzi C.,Chincarini G., Pasotti F.,Covino S., Mao J., 2008, ArXiv:0809.0189
\bibitem[Marisaldi et~al.(2008)]{Marisaldi08} Marisaldi M., et al., 2008, GCN circular, 7457
\bibitem[M\'esz\'aros (2002)]{Mesz02}M\'esz\'aros P., 2002, ARA\&A, 40, 137
\bibitem[M\'{e}sz\'{a}ros \& Rees (1994)]{MR94} M\'{e}sz\'{a}ros P., Rees M. J., 1994, MNRAS, 269, L41
\bibitem[Mimica et al. (2008)]{Mimica08}Mimica P., Giannios D., Aloy M. A., 2008, A\&A, in press (arXiv:0810.2961)
\bibitem[Nakar \& Piran (2004)]{NP04} Nakar E., Piran T., 2004, MNRAS, 353, 647
\bibitem[Nakar \& Piran (2005)]{NP05} Nakar E., Piran T., 2005, ApJ, 619, L147
\bibitem[Omodei et~al.(2008)]{Omodei08} Omodei N., et~al., 2008, GCN circular, 8407
\bibitem[Omodei (2008)]{Omodei08b} Omodei N., 2008, talk given at the ``6th Workshop on Science with the New Generation of High Energy Gamma-ray Experiments"
(http://glast.pi.infn.it/presentations/081008\_GRB\_Abano.ppt)
\bibitem[Panaitescu \& Kumar (2001)]{PK01} Panaitescu A.,
Kumar P., 2001, ApJ, 554, 667
\bibitem[Piran (2004)]{Piran04} Piran T., 2004, Rev. Mod. Phys., 76, 1143
\bibitem[Piran, Sari \& Zou(2008)]{psz08}Piran T., Sari R.,  Zou Y. C., 2008, MNRAS, in press
(arXiv:0807.3954)
\bibitem[Racusin et al.(2008b)]{Racusin08b} Racusin J., et al., 2008b, GCN circular, 7427
\bibitem[Racusin et al.(2008)]{Racusin08}Racusin J. L., et al., 2008, Nature, 455, 183
\bibitem[Sari \& Esin(2001)]{se01}Sari, R., Esin, A., 2001, ApJ, 548, 787
\bibitem[Sari et al. (1998)]{SPN98} Sari R., Piran T.,  Narayan R.,
1998, ApJ, 497, L17
\bibitem[Stecker, Malkan \&  Scully(2006)]{stecker06} Stecker F. W., Malkan, M. A.,  Scully S. T., Astrophys. J., 2006, 648, 774
\bibitem[Tajima et~al.(2008)]{Tajima08}Tajima H., et~al., 2008, GCN circular, 8246
\bibitem[van der Horst \& Goldstein (2008)]{van08} van der Horst A., Goldstein A.,
2008, GCN circular, 8278
\bibitem[Vreeswijk et al.(2008)]{Vreeswijk08} Vreeswijk P. M., et al., 2008, GCN circular, 7444
\bibitem[Wang et al. (2001)]{wdl01}Wang X. Y., Dai Z. G. \& Lu T., 2001, ApJ, 556, 1010
\bibitem[Wang et al. (2006)]{wlm06}Wang X. Y., Li Z., M\'esz\'aros P., 2006, ApJ, 641,
L89
\bibitem[Wang \& M\'esz\'aros(2006)]{wp06}Wang X. Y., M\'esz\'aros P., 2006, ApJ, 643, L95
\bibitem[Wei et al. (2006)]{Wei06} Wei D. M., Yan T., Fan Y. Z., 2006, ApJ, 636, L69
\bibitem[Wu et al. (2008)]{Wu08}Wu X. F., et al., 2008, ApJ, to be submitted
\bibitem[Zhang (2006)]{Zhang06} Zhang B., 2006, AIPC, 838, 392
\bibitem[Zhang (2007)]{Zhang07} Zhang B., 2007, ChJAA, 7, 1
\bibitem[Zhang et al. (2003)]{zhang03} Zhang B., Kobayashi S., M\'esz\'aros P. 2003, ApJ,
595, 950
\bibitem[Zhang \& M\'esz\'aros(2001)]{zm01}Zhang B., M\'esz\'aros P., 2001, ApJ, 559, 110
\bibitem[Zou, Wu \& Dai(2005)]{zwd05} Zou Y. C., Wu X. F., Dai Z. G., 2005, MNRAS, 363, 93
\bibitem[Zou, Piran \& Sari(2009)]{zps08}Zou Y. C., Piran T., Sari R., 2009,
ApJ, 692, L92

\end{thebibliography}
\end{document}